\documentclass[10pt]{article}
\usepackage{amsfonts}
\usepackage{amsmath}
\usepackage{amssymb}
\usepackage{graphicx}
\usepackage{color}

\def \be {\begin{equation}}
\def \ee {\end{equation}}
\def \bea {\begin{eqnarray}}
\def \eea {\end{eqnarray}}
\def \nn {\nonumber}
\def \la {\langle}
\def \ra {\rangle}
\def \rr {\raise.35ex\hbox{\small $\prime$}\kern-.17em{\mbox{\large $\imath$}}}
\def \del {\partial}
\def \dels {\partial\kern-.5em / \kern.5em}
\def \As {{A\kern-.5em / \kern.5em}}
\def \Ds {D\kern-.7em / \kern.5em}

\def \a {\alpha}

\def \d {\delta}
\def \eps {\epsilon}

\def \lam {\lambda}

\def \s {\sigma}

\def \II {I\hspace{-.1em}I\hspace{.1em}}

\def \D {{\mathcal{D}}}

\newcommand{\ba}{\begin{eqnarray}}
\newcommand{\ea}{\end{eqnarray}}
\newcommand{\cA}{\mathcal{A}}
\newcommand{\cH}{\mathcal{H}}

\newcommand{\cK}{\mathcal{K}}
\newcommand{\cJ}{\mathcal{J}}
\newcommand{\cZ}{\mathcal{Z}}
\newcommand{\bi}{{\bar \imath}}
\newcommand{\bj}{{\bar \jmath}}
\newcommand{\bk}{\bar k}
\newcommand{\bl}{\bar l}
\newcommand{\bm}{\bar m}

\newcommand{\e}{\mathbf{e}}
\newcommand{\vv}{\mathbf{v}}
\newcommand{\psib}{{\bar \psi}}

\begin{document}
\begin{titlepage}
%\catcode`\@=11
%\catcode`\@=12
%\twocolumn[\hsize\textwidth\columnwidth\hsize\csname%
%@twocolumnfalse\endcsname

%\draft
\begin{center}
%\hfill hep-th/yymmnnn\\
\hfill UT-08-08
\vskip .5in

\textbf{\LARGE Lie 3-Algebra and Multiple M2-branes}

\vskip .5in
{\large
Pei-Ming Ho$^\dagger$\footnote{
e-mail address: pmho@phys.ntu.edu.tw}, 
Ru-Chuen Hou$^\dagger$,
Yutaka Matsuo$^\ddagger$\footnote{
e-mail address:
 matsuo@phys.s.u-tokyo.ac.jp}}\\
\vskip 3mm
{\it\large
$^\dagger$
Department of Physics and Center for Theoretical Sciences, \\
National Taiwan University, Taipei 10617, Taiwan,
R.O.C.}\\
\vskip 3mm
{\it\large
$^\ddagger$
Department of Physics, Faculty of Science, University of Tokyo,\\
Hongo 7-3-1, Bunkyo-ku, Tokyo 113-0033, Japan\\
\noindent{ \smallskip }\\
}
\vspace{60pt}
%\maketitle
\end{center}
\begin{abstract}
Motivated by the recent proposal of 
an $N=8$ supersymmetric action 
for multiple M2-branes,
we study the Lie $3$-algebra in detail. 
In particular, we focus on the fundamental identity 
and the relation with Nambu-Poisson bracket. 
Some new algebras not known in the literature are found. 
Next we consider cubic matrix representations 
of Lie 3-algebras. 
We show how to obtain higher dimensional
representations by tensor products for a generic 3-algebra. 
A criterion of reducibility is presented.  
We also discuss the application of Lie 3-algebra 
to the membrane physics, 
including the Basu-Harvey equation and 
the Bagger-Lambert model. 

\end{abstract}

%\pacs{PACS numbers: 11.25.-w, 11.25.Mj, 11.25.Sq}%]
\end{titlepage}
%\begin{narrowtext}
\setcounter{footnote}{0}

\section{Introduction}

In the long history of the study of Nambu bracket 
\cite{Nambu}, the relation with the supermembrane or
M-theory has been giving the main motivation
(see \cite{Berman:2007bv} for the references). 
There have been many attempts
to quantize the classical Nambu bracket toward this direction.  
However,  since the quantization is difficult and does not seem to be unique, 
we need to understand which properties are essential from the
physical viewpoint.

Recently Bagger and Lambert \cite{Bagger:2006sk,Bagger:2007jr,Bagger:2007vi}
and Gustavsson \cite{Gustavsson:2007vu} proposed
a formalism of multiple M2-branes and it was found that
the generalized Jacobi identity (or the fundamental identity)
for Lie $3$-algebra is essential to define the action with
$\mathcal{N}=8$ supersymmetry.
It seems to give the desired principle of 
constructing quantum Nambu bracket which has been long sought for. 
So far the only explicit example of Lie 3-algebra 
ever considered for the Bagger-Lambert model is ${\cal A}_4$, 
the $SO(4)$-invariant algebra with 4 generators.\footnote{
See also Kawamura's work \cite{Kawamura:2002yz,Kawamura:2003cw}
where the same algebra and its representation was studied.
}
For a more concrete understanding of the Bagger-Lambert model, 
it is urgent to study more explicit examples of Lie 3-algebra. 
In the mathematical literature, the Lie $3$-algebra 
(also known as Filippov algebra) is not new \cite{Filippov}, 
and its structure has been studied to some extent.
However, not only that the complete classification of the algebra
does not exist, there are very few explicit examples in the literature.  

In this paper, we first endeavor to find new examples 
of Lie 3-algebra 
(section \ref{sec:Filippov}).
After a survey of the mathematical literature, 
especially the study of Nambu-Poisson bracket, 
interestingly, we successfully find several new examples
(section \ref{sec:examples}). 
All the new examples have one important feature in common, 
namely that their metrics are not positive-definite. 
In this respect they are very different from $\cA_4$. 
We also tried to search for solutions of the fundamental identity 
with positive-definite metrics by computer when the number of generators
are small ($n=5,6,7,8$), and found that there are no
algebras except for ${\cal A}_4$ and its direct sum. 
We are led to make the conjecture that 
there are no other 3-algebras with a positive definite metric. 
Generators of zero norm are almost ubiquitous in 3-algebras. 

In section \ref{sec:representation}, 
we consider the problem of realizing Lie 3-algebras 
using cubic matrices. 
As an example, we consider cubic-matrix representations
for ${\cal A}_4$, 
and try to develop a systematic method to generate 
higher dimensional representations. 
In the case of Lie algebra, a simple method to derive 
higher dimensional representations is to use the tensor product 
and then to decompose it into irreducible representations.
Here we show that we can do similar construction of
higher dimensional representations by tensor product.
One can define the notion of irreducibility similarly, 
although we need to redefine the product of cubic matrices.  

In section \ref{sec:application}, 
we review Basu-Harvey equation, 
and demonstrate that its success in describing 
the configuration of multiple M2-branes ending on an M5-brane 
does not reply on the specific realization of the 3-algebra 
as it was originally considered. 
We only need the 3-algebra structure for the calculation. 
We also comment on its relation to the Bagger-Lambert model. 
A few comments about future directions are made
 in section \ref{sec:comments}. 

In appendix A, we point out the relation between the fundamental
identity and the Pl\"ucker relation.
The latter appeared frequently
in the literature of the exactly solvable system, matrix model
and topological strings.

\section{Lie $n$-Algebra}\label{sec:Filippov}

\subsection{Definitions}
\label{sec:Definition}

Lie $n$-algebra, also known as $n$-ary Lie algebra, or Filippov $n$-algebra 
\cite{Filippov}, 
is a natural generalization of Lie algebra. 
For a linear space 
${\cal V} = \{ \sum_{a=1}^{\cal D} v_a T_a ; v_a \in \mathbb{C} \}$
of dimension ${\cal D}$, 
a Lie $n$-algebra structure is defined by 
a multilinear map called Nambu bracket 
$[ \cdot, \cdots, \cdot ]: {\cal V}^{\otimes n} \rightarrow {\cal V}$ 
satisfying the following properties 
\footnote{
In part of the literature \cite{naryLie2}, 
the fundamental identity (\ref{FI}) 
is replaced by a weaker (skew-symmetrized) version, 
and thus the definition of Lie $n$-algebra is ambiguous. 
The definition we consider here is more closely related to 
the physical applications we will consider below. 
See also \cite{Curtright:2002fd} for various aspects of the classical
and quantum Nambu bracket.
}
\begin{enumerate}
\item
Skew-symmetry:
\be
[A_{\s(1)}, \cdots, A_{\s(n)}] = (-1)^{|\s|} [A_1, \cdots, A_n].
\ee
\item
Fundamental identity:
\be \label{FI} 
[A_1, \cdots, A_{n-1}, [B_1, \cdots, B_n]]
= \sum_{k=1}^n [B_1, \cdots, B_{k-1}, [A_1, \cdots, A_{n-1}, B_k ], 
B_{k+1}, \cdots, B_n].
\ee
\end{enumerate}
The fundamental identity is also called 
the generalized Jacobi identity. 
It means that the bracket $[A_1, \cdots, A_{n-1}, \cdot]$ 
acts as a derivative on ${\cal V}$, 
and it may be used to represent a symmetry transformation. 

In terms of the basis, $n$-algebra is expressed in terms of the
(generalized) structure constants,
\ba
&& [T_{a_1},\cdots, T_{a_n}] = i {f_{a_1 \cdots a_n}}^b \,T_b
\ea
The  fundamental identity implies a bilinear relation 
the structure constants,
\ba
&& \sum_{c} f_{b_1 \cdots b_p}{}^c f_{a_1 \cdots a_{p-1} c}{}^{d}=
\sum_i\sum_c {f_{a_1 \cdots a_{p-1} b_i}}^c 
{f_{b_1 \cdots c \cdots b_p}}^d\,. \label{fip}
\ea

One may introduce the inner product in the space of algebra $\cA$
as a bilinear map from ${\cal V}\times {\cal V}$ to $\mathbb{C}$ 
\ba
\la T_a, T_b \ra =h_{ab}\,.
\ea
We will refer to the symmetric tensor $h_{ab}$ as the metric in the following.
As a generalization of the Killing form in Lie algebra,
we require that the metric is invariant under any transformation 
generated by the bracket $[T_{a_1}, \cdots, T_{a_{n-1}}, \cdot ]$: 
\ba
\la [T_{a_1},\cdots, T_{a_{n-1}}, T_b], T_c \ra 
+ \la T_b,[T_{a_1}, \cdots, T_{a_{n-1}}, T_c] \ra = 0. 
\ea
This implies a relation for the structure constant 
\ba\label{invmetric}
h_{cd}{f_{a_1\cdots a_{n-1}b}}^d+h_{bd}{f_{a_1\cdots a_{n-1}c}}^d = 0\,.
\ea
Therefore the tensor 
\be \label{tat}
f_{a_1 \cdots a_n} \equiv f_{a_1 \cdots a_{n-1}}{}^b h_{b a_n} 
\ee
is totally antisymmetrized. 

For applications to physics, it is very important to 
have a nontrivial metric $h_{ab}$ in order to write 
down a Lagrangian or physical observables which 
are invariant under transformations defined by 
$n$-brackets. 

Another mathematical structure of physical importance is 
Hermitian conjugation. 
A natural definition of the Hermitian conjugate of an $n$-bracket is 
\be \label{Hermiticity}
[A_1, \cdots, A_n]^{\dagger} = [A_n^{\dagger}, \cdots, A_1^{\dagger}].
\ee
This relation determines the reality of structure constants. 
For the usual Lie algebra, 
if we choose the generators to be Hermitian, 
the structure constants $f_{ab}{}^c$ are real numbers, 
and if the generators are anti-Hermitian, 
the structure constants are imaginary. 
This is not the case for $3$-brackets. 
The structure constants are always imaginary 
when the generators are all Hermitian or all anti-Hermitian. 
In general, for $n$-brackets, 
the structure constants are real if $n = 0,1$ (mod 4), 
and imaginary if $n = 2, 3$ (mod 4) 
for Hermitian generators. 
The structure constants are multiplied by a factor of $\pm i$ 
when we replace Hermitian generators by 
anti-Hermitian ones only for even $n$.  

From now on we will focus on the case of $n=3$. 
Explicitly, for $3$-algebra the fundamental identity (\ref{FI}) is
\be
[A_1, A_2, [B_1, B_2, B_3]] = 
[[A_1, A_2, B_1], B_2, B_3] + [B_1, [A_1, A_2, B_2], B_3]
+[B_1, B_2, [A_1, A_2, B_3]].
\ee
In terms of the structure constant, 
the fundamental identity is 
\ba\label{fi}
\sum_{i} {f_{cde}}^i {f_{abi}}^j=\sum_i\left(
{f_{abc}}^i {f_{ide}}^j+{f_{abd}}^i {f_{cie}}^j+{f_{abe}}^i {f_{cdi}}^j
\right)\,.
\ea

One of the important questions is how to classify the solutions of
the fundamental identity (\ref{fi}) (or more generally (\ref{fip})).
The trivial solution is to put all structure constants zero ${f_{abc}}^d=0$.
The simplest nontrivial solution which satisfy the fundamental identity
(\ref{fi}) of $3$-algebra starts from $\D=4$, 
\be \label{algebra}
[T_a, T_b, T_c] = i \eps_{abcd} T_d, \qquad (a,b,c,d=1,2,3,4), 
\ee
and the metric is fixed by the requirement of invariance (\ref{invmetric}) to be 
\be \label{deltametric} 
h_{ab} = \delta_{ab} 
\ee
up to an overall constant factor. 
Compared with the formula in some literature, 
we have an extra factor of $i$ on the right hand side of (\ref{algebra}) 
due to our convention of the Nambu bracket's Hermiticity (\ref{Hermiticity}).

This algebra is invariant under  $SO(4)$,
and will be denoted as $\cA_4$.
The structure constant is given by
the totally antisymmetrized epsilon tensor  
$
{f_{abc}}^d=i\,\epsilon_{abcd}
$. 
In general, for any $n$, the fundamental identity (\ref{fip}) is solved by
the epsilon tensor in $\D=n+1$,
\ba
{f_{a_1\cdots a_n}}^b=i\,\epsilon_{a_1\cdots a_n b}\, ,
\ea
with the metric (\ref{deltametric}). 

From these algebras, one may obtain higher rank algebras by direct sum
as usual. For $n=3$ case, the
algebra $\cA_4\oplus\cdots \oplus \cA_4$ ($p$-times) with $\D=4p$
is written as,
\ba
&& [T_a^{(\alpha)}, T_{b}^{(\beta)}, T_{c}^{(\gamma)}] =
 i \eps_{abcd} \delta_{\alpha\beta\gamma\delta}
T_{d}^{(\delta)}, \\
&& ~~~~~~ (a,b,c,d=1,2,3,4,\quad \alpha,\beta,\gamma,\delta=1,\cdots, p)\,,\nn
\ea
where $\delta_{\alpha\beta\gamma\delta}=\delta_{\alpha\beta}\delta_{\alpha\gamma}
\delta_{\alpha\delta}$.

A nontrivial question is whether there exists any 3-algebra
which can not be reduced to the direct sums of the algebra $\cA_4$, 
up to a direct sum with a trivial algebra. 
For $n=3$, one may directly solve the
fundamental identity by computer 
for lower dimensions $\D$. 
We have examined the cases $\D=5,6,7,8$ 
with the assumption that the metric $h_{ab}$
is invertible and can be set to $\delta_{ab}$ after the change of basis. 
In this case the structure constant ${f_{abc}}^d$ can be identified
with totally anti-symmetric four tensor $f_{abcd}$.

For $\D=5,6$, one can solve directly the fundamental
identity algebraically  by computer. 
For $\D=7,8$, we assume the coefficients $f_{abcd}$
are integer and $|f_{abcd}|\leq 3$ and scanned all possible combinations.
After all, the solutions can always be reduced to $\cA_4$ 
up to a direct sum with a trivial algebra, 
or $\cA_4\oplus \cA_4$ ($\D=8$) after a change of basis.\footnote{
One of the failed examples is,
\ba
&&\sum_{a,b,c,d=1}^7 f_{abcd}\e_a\wedge \e_b\wedge \e_c\wedge
\e_d = \e_1\wedge \e_2\wedge \e_3\wedge \e_4+ \e_1 \wedge \e_2
\wedge \e_5 \wedge \e_6 -\e_1\wedge \e_3\wedge \e_5 \wedge \e_7\nn\\
&&~~~~~~~~~+\e_1 \wedge \e_4 \wedge \e_6 \wedge \e_7 +
\e_2 \wedge \e_3 \wedge \e_6 \wedge \e_7
+\e_2 \wedge \e_4 \wedge \e_5 \wedge \e_7 +
\e_3\wedge \e_4\wedge \e_5 \wedge \e_6
\ea
for $\D=7$. This is the Hodge dual of $G_2$-invariant 3-form.
It was also mentioned in \cite{Bandres:2008vf}.}
This observation suggests that the Lie $n$-algebra for $n>2$ is
very limited.  

Actually there is an interesting relation between the fundamental identity
and the Pl\"ucker  relation (for the Grassmaniann manifold), which
 will be explained in the appendix.
It automatically tells us that the epsilon tensor is the solution 
of the fundamental identity for Lie $n$-algebra in general.
At the same time, it also implies that to find other solutions
are very difficult.

While very little is known about explicit nontrivial examples 
of the $n$-algebra, 
its correspondence with Nambu-Poisson brackets 
given in \S \ref{sec:NP} is very helpful.

If the metric is not invertible, it becomes possible to construct
Lie $3$-algebra other than the direct sum of $\cA_4$.  We will
construct some examples in \S \ref{sec:examples}.

\subsection{Review of Nambu-Poisson Brackets} 
\label{sec:NP}

Let ${\cal M}_d$ be a manifold of $d$ dimensions, 
and $C({\cal M}_d)$ its algebra of functions. 
A Nambu-Poisson bracket 
is a multi-linear map from $C({\cal M}_d)^{\otimes 3}$ to $C({\cal M}_d)$ 
that satisfies the following conditions \cite{Takhtajan}:
\begin{enumerate}
\item
Skew-symmetry:
\be
\{f_{\s(1)}, f_{\s(2)}, f_{\s(3)}\} = (-1)^{|\s|} \{f_1, f_2, f_3\}.
\ee
\item
Leibniz rule:
\be
\{f_1, f_2, gh\} = \{f_1, f_2, g\}h + 
g\{f_1, f_2, h\}.
\ee
\item
Fundamental identity:
\be\label{fi:NP}
\{g, h, \{f_1, f_2, f_3\}\} = 
\{\{g, h, f_1\}, f_2, f_3\} + \{f_1, \{g, h, f_2\}, f_3\}
+\{f_1, f_2, \{g, h, f_3\}\}.
\ee
\end{enumerate}

The prototype of a Nambu-Poisson bracket is 
the Jacobian determinant for 3 variables $x_i (i = 1, 2, 3)$
\be
\{ f_1, f_2, f_3 \} = \eps_{ijk} \del_i f_1 \del_j f_2 \del_k f_3.
\ee
where $i, j, k = 1, 2, 3$. This is the classical Nambu bracket.
More general Nambu-Poisson bracket can be written
in terms of the local coordinates as,
\ba
\{ f_1, f_2, f_3 \} = \sum_{i_1<i_2< i_3}\sum_{\sigma\in S_3}
(-1)^\sigma P_{i_1 i_2 i_3}(x)\del_{i_{\sigma(1)}} f_1 \del_{i_{\sigma(2)}} f_2 
\del_{i_{\sigma(3)}}  f_3.
\ea

It is proved that one can always choose 
coordinates such that any Nambu-Poisson bracket 
is locally just a Jacobian determinant \cite{Jacobian}. 
Locally we can choose coordinates such that 
\be
\{f, g, h\} = \eps^{ijk}\del_i f\; \del_j g\; \del_k h, 
\ee
where $i, j, k = 1,2,3$,  
and $dx_1 dx_2 dx_3$ defines a local expression of the volume form. 
As a result, it is straightforward to check that 
the Nambu-Poisson bracket can be used to 
generate volume-preserving diffeomorphisms on a function $f$ 
\be
\d f = \{ g_1, g_2, f \} 
\ee
specified by two functions $g_1$ and $g_2$. 

A Nambu-Poisson algebra is also an infinite dimensional Lie 3-algebra. 
For a 3-manifold on which the Nambu-Poisson bracket is everywhere 
non-vanishing, 
it is natural to use the volume form picked by the bracket 
to define an integral $\int_{\cal M}$, 
and then the metric can be defined by 
\be
\la f, g\ra = \int_{\cal M} fg.
\ee
Symmetries of the algebra are then automatically preserved by the metric. 
% {\em (YM: Is it true and in what sense and in what conditions? 
% The direct sum algebra is a
% counter example of this statement.
% PM: Did you mean a direct sum given by, say, 
% $P = \del_1\wedge \del_2\wedge \del_3 + \del_4\wedge \del_5\wedge \del_6$? 
% This one is not consistent. 
% Consider, for instance the FI for 
% \[
% (x_1 x_4, x_2, (x_3, x_5, x_6)) = \cdots. 
% \]
% You get inconsistency.  
% That statement I made seems to be what people say in the literature. 
% See, e.g. Theorem 2.2 on p. 218 in \cite{Vaisman}, 
% or the comment at the bottom of p.2 in \cite{GrabowskiMarmo}.)}

The notion of Nambu-Poisson brackets can be 
naturally generalized to brackets of order $n$, 
as a map from $C({\cal M}_d)^{\otimes n}$ to $C({\cal M}_d)$.
The fundamental identity for Nambu-Poisson brackets of order $n$ is
\be
\{f_1, \cdots, f_{n-1}, \{g_1, \cdots, g_n\}\}
= \sum_{k=1}^n \{g_1, \cdots, g_{k-1}, \{f_1, \cdots, f_{n-1}, g_k \}, 
g_{k+1}, \cdots, g_n\}.
\ee
Both the Leibniz rule and the fundamental identity 
indicate that it is natural to think of
\be
\{f_1, \cdots, f_{n-1}, \; \cdot \; \}: C({\cal M}_d) 
\rightarrow C({\cal M}_d) 
\ee
as a derivative on functions.

Each Nambu-Poisson bracket of order $n$ corresponds to 
a Nambu-Poisson tensor field $P$ through the relation
\ba
&&\{f_1, \cdots, f_n\} = P(df_1, \cdots, df_n),\\
&& P=\sum_{i_1<\cdots< i_n} 
P_{i_1\cdots i_n}(x)\partial_{i_1}\wedge\cdots 
\wedge \partial_{i_n}\,.
\ea
The theorem mentioned above can also be generalized 
to brackets of order $n$, 
which means that any Nambu-Poisson tensor field $P$ 
is decomposable, i.e., 
one can express $P$ as
\be
P = V_1 \wedge \cdots \wedge V_n
\ee
for $n$-vector fields $V_i$.
% {\em (YM: This statement should be made more carefully since
% it seems to contradict the theorem for the linear case)}
For a review of Nambu-Poisson brackets see, e.g. \cite{Vaisman}.

Let us now focus on the case $n=3$. 
When all the coefficients of the Nambu-Poisson tensor field are linear in $x$, 
that is, $P_{i_1i_2 i_3}(x)=\sum_j {f_{i_1 i_2 i_3}}^j x_j$ 
for constant $f_{i_1 i_2 i_3}{}^j$, 
we call the bracket a linear Nambu-Poisson bracket, 
and it takes the form of a Lie $3$-algebra on the coordinates 
\ba
\{x_i, x_j, x_k\}=\sum_l {f_{ijk}}^l x_l. 
\ea
Apparently,  
a linear Nambu-Poisson bracket is also a Lie $3$-algebra 
when we restrict ourselves to linear functions of the coordinates $x_i$.
We have to be careful, however, in that 
the reverse is not true, as they also have some differences.
For the Nambu-Poisson bracket, one may change the coordinates by
a general coordinate transformation.  
On the other hand, for Lie $3$-algebra, we
only allow linear transformations of the basis.  
Since the requirement of Leibniz rule for the Nambu-Poisson bracket 
is not imposed on a Lie 3-algebra, 
we expect that only a small fraction of Lie 3-algebras are 
also linear Nambu-Poisson algebras. 
In particular, we do not expect that the Nambu bracket 
of a generic Lie 3-algebra be decomposable. 

It has been shown that any linear Nambu-Poisson tensor of order $n$ 
on a linear space $V_d$ can be put in 
one of the following forms 
by choosing a suitable basis of $V_d$ \cite{DZ}: 
\begin{enumerate}
\item Type I:
\be \label{type1-1}
P_{(r,s)} = \sum_{j=1}^{r+1} \pm x_j \del_1\wedge\cdots
\wedge\del_{j-1}\wedge\del_{j+1}\wedge
\cdots\wedge\del_{n+1}
+\sum_{j=1}^s \pm x_{n+j+1}\del_1\wedge\cdots
\wedge\del_{r+j}\wedge\del_{r+j+2}\wedge\cdots
\wedge\del_{n+1},
\ee
where $-1\leq r\leq n$, $0\leq s\leq \mbox{min}(d-n-1,n-r)$.
Explicitly, we have 
\be \label{type1-2}
\{x_1, \cdots, x_{j-1}, x_j, \cdots, x_{n+1} \} =
\left\{
\begin{array}{ll}
\pm x_j, & 1 \leq j \leq r+1, \\
\pm x_{j-r+3}, & r+2 \leq j \leq r+s+1, \\
0, & r+s+2 \leq j \leq d. 
\end{array}
\right.
\ee
\item Type \II:
\be \label{type2-1}
P = \del_1\wedge\cdots\wedge\del_{n-1}\wedge
\left(\sum_{i,j=n}^d a_{ij} x_i\del_j\right).
\ee
In other words, 
\be \label{type2-2}
\{ x_1, \cdots, x_{n-1}, x_j \} = \sum_{i=n}^d a_{ij} x_i, 
\qquad
j=n, \cdots, d. 
\ee
\end{enumerate}

Here the choice of coordinates is made such 
that the Nambu-Poisson tensor field is linear, 
instead of trying to make its decomposability manifest. 
When we interpret these brackets as Nambu brackets 
on the linear space generated by $\{ x_i \}$, 
we are no longer allowed to make general coordinate 
transformations on the generators $x_i$, 
and the decomposability of the Nambu-Poisson tensor field 
is no longer relevant.

\section{Examples of Lie $3$-Algebra}
\label{sec:examples}

We already know a few examples of Lie $3$-algebra
which satisfies the fundamental identity.
\begin{itemize}
\item
A trivial algebra is one for which the Nambu bracket is always 0. 

\item
The 4-generator algebra 
with $SO(4)$ symmetry $\cA_4$.
% \be
% [ x_i, x_j, x_k ] = \epsilon_{ijkl} x_l, \qquad
% i,j,k,l=1,2,3,4,
% \ee
% will be denoted as ${\cal A}_4$.

\item Direct sums of an arbitrary number of copies of $\cA_4$ and a trivial algebra.

\item
All Nambu-Poisson brackets on $C({\cal M}_d)$ 
are of course also Nambu brackets on 
the infinite dimensional linear space $C({\cal M}_d)$. 
\end{itemize}

In the following, we list a few more examples of Lie $3$-algebra. 
In contrast with previous studies on this problem, 
we put relatively more emphasis on the metric, 
which is crucial for writing down an invariant observable or Lagrangian. 
\footnote{
However,  \cite{Gran:2008vi} 
suggests that we study the Bagger-Lambert model 
only at the level of equations of motion, 
which can be described without a metric. 
}
Besides ${\cal A}_4$, the only well known example of 3-algebra 
is the class constructed in \cite{Awata:1999dz}. 
However, as we will show below in section \ref{onegenerator}, 
the invariant metric is almost trivial in those cases.

\subsection{Linear Nambu-Poisson Bracket: Type I }

First, since any linear Nambu-Poisson bracket is also a Lie $3$-algebra,
the classification of the last subsection gives type I and type \II algebras. 

A type I linear Nambu-Poisson bracket $P_{(r, s)}$ 
(\ref{type1-1}, \ref{type1-2})
is labeled by a pair of integers $(r, s)$. 
$P_{(3, 0)}$ in (\ref{type1-1}) with plus signs 
for $n = 3$ gives $\cA_4$ algebra.
For other values of $(r,s)$, 
$P_{(r,s)}$ gives a new algebra. 

For example, $P_{(-1,4)}$
defines an algebra with 8 generators 
(apart from direct sum with a trivial algebra) 
\ba
[T_2,T_3, T_4]= \pm T_5,\quad
[T_1,T_3, T_4]= \pm T_6,\quad
[T_1,T_2, T_4]= \pm T_7,\quad
[T_1,T_2,T_3]= \pm T_8.\quad
\ea
Without loss of generality, 
we can take all plus signs above, 
and an invariant metric is given by 
\be
h_{15} = - h_{26} = h_{37} = - h_{48} = K 
\ee
for some constant $K$.
The metric is thus non-degenerate 
with the signature $(++++----)$. 

Another example is $P_{(1,1)}$, which is defined by 
\be
[T_2, T_3, T_4] = -T_1, \quad
[T_1, T_3, T_4] = \eps T_2, \quad
[T_1, T_2, T_4] = T_5, \quad
[T_1, T_2, T_3] = T_6, 
\ee 
where we have fixed the signs except $\eps = \pm 1$ 
by convention. 
The invariant metric is given by 
\be
h_{11} = \eps h_{22} = h_{35} = - h_{46} = 1, 
\ee
while other components of $h$ vanish.

\subsection{Linear Nambu-Poisson Bracket: Type \II }

The linear Nambu-Poisson algebra of type \II (\ref{type2-1}, \ref{type2-2})
for arbitrary constant matrix $a_{ij}$ has the Nambu bracket 
\ba
[T_1,T_2, T_j]= \sum_{i=3}^d a_{ij} T_i \quad
(j=3,\cdots, d)\,.
\ea
The invariance of the metric implies that
\be
h_{i1} = h_{i2} = \sum_{i=3}^d h_{ji}a_{ik} = 0 
\ee
for $i, j, k = 3, \cdots, d$. 
Thus $a = 0$ if $h$ is invertible. 
Conversely, if $a$ is invertible then $h_{ij} = 0$ 
for $i, j = 3, \cdots, d$. 
As $T_1$ and $T_2$ do not appear on 
the right hand side of the Nambu bracket, 
there is no constraint on $h_{11}, h_{12}$ or $h_{22}$. 

As Nambu-Poisson brackets, 
we can extend the 3-algebra on the space 
of linear functions ${\cal V} = \{ \sum_{i=1}^d a_i T_i \}$ 
to all polynomials of $T_i$'s. 
The product of $T_i$'s defines a commutative algebra.

\subsection{One-Generator Extension of a Lie Algebra}
\label{onegenerator}

In addition, we may construct other examples. 
For a given Lie algebra ${\cal G}$
with generators $T_a$ and structure constants $f_{ab}{}^c$,
we can introduce a new element $T_0$ and
define a Lie 3-algebra by \cite{GrabowskiMarmo} 
\bea 
[T_0, T_a, T_b] &=& f_{ab}{}^c T_c, \label{T0Ta1} \\
{[T_a, T_b, T_c ]} &=& 0 \label{T0Ta2}
\eea
for $a, b, c = 1, \cdots, \mbox{dim}\;{\cal G}$. 
For a simple Lie algebra ${\cal G}$, 
the invariance of the metric demands that
\be
\la [T_0, T_a, T_b], T_c\ra + \la T_b, [T_0, T_a, T_c]\ra = 0 
\quad \Rightarrow \quad 
f_{ab}{}^d h_{dc} + f_{ac}{}^d h_{db} = 0. 
\ee
This suggests that $h_{ab}$ should be proportional 
to the Killing form of ${\cal G}$. 
However, the invariance conditions also include 
\bea
\la [T_a, T_b, T_c], T_0\ra + \la T_c, [T_a, T_b, T_0]\ra = 0 
&\Rightarrow& f_{ab}{}^d h_{dc} = 0, \nn \\
\la [T_a, T_b, T_0], T_0 \ra + \la T_0, [T_a, T_b, T_0]\ra = 0 
&\Rightarrow& h_{c0} = 0. 
\eea
Therefore, we can not use the Killing form 
of the Lie algebra ${\cal G}$ as $h_{ab}$, 
but instead the metric should be taken as 
\be
h_{ab} = h_{0a} = 0, 
\qquad 
h_{00} = K, \qquad 
a, b = 1, \cdots, \mbox{dim}\;{\cal G}, 
\ee
where $K$ is an arbitrary constant. 

If the Lie algebra ${\cal G}$ can be realized as a matrix algebra, 
this 3-algebra can also be extended to polynomials of $T_a$'s. 
(That is, we extend the Lie algebra ${\cal G}$ 
to its universal enveloping algebra.) 
We can define the Nambu bracket by 
\be
[T_0, A, B] = [A, B] \equiv AB-BA, \qquad 
[A, B, C] = 0, 
\ee
where $A, B, C$ are elements of the matrix algebra. 
The Leibniz rule follows from this definition 
\footnote{
Note that here the ordering of the product on the right hand side 
is important, unlike the case of a Nambu-Poisson algebra. 
} 
\be
[T_0, A, BC] = [T_0, A, B] C + B[T_0, A, C].
\ee
However, it is not possible for the Leibniz rule to
apply to products involving $T_0$.  

This 3-algebra has a close connection with 
the Nambu bracket defined in \cite{Awata:1999dz}. 
For a matrix algebra, the Nambu bracket in \cite{Awata:1999dz}
is defined as 
\be
[A, B, C] = \mbox{tr}(A)[B, C] + \mbox{tr}(B)[C, A] + \mbox{tr}(C)[A, B].
\ee
This Nambu bracket is automatically skew-symmetric 
and satisfies the fundamental identity. 
For a matrix algebra, we can choose the basis of generators 
such that there is only one generator, the identity $I$, 
that has a non-vanishing trace. 
Denoting $T_0 = I/\mbox{tr}(I)$, and 
the rest of the generators as $T_a$ $(a\neq 0)$, 
the Nambu bracket is precisely given by (\ref{T0Ta1}) and (\ref{T0Ta2}). 
Thus we see that the Nambu bracket of \cite{Awata:1999dz} 
is equivalent to the 3-algebra in this subsection 
for the case when ${\cal G}$ is a matrix algebra of traceless matrices.

\subsection{A Truncation of Nambu-Poisson Structure on $S^3$}
\label{truncateS3}

The classical Nambu bracket 
\be \label{classN}
\{ f_1, f_2, f_3 \} = x_i \; \epsilon_{ijkl} 
\; \del_j f_1 \; \del_k f_2 \; \del_l f_3  
\ee
defines a Nambu-Poisson bracket with $SO(4)$ symmetry
on the space of all polynomials of $\{x_i: i = 1,\cdots, 4\}$ to all order. 
Based on this we define a Nambu bracket 
which is restricted to polynomials of order no larger than $N$ as
\be \label{Ntrunc}
[ X_{i_1\cdots i_l}, X_{j_1\cdots j_m}, X_{k_1\cdots k_n} ] 
= \left\{ 
\begin{array}{ll}
\{ X_{i_1\cdots i_l}, X_{j_1\cdots j_m}, X_{k_1\cdots k_n} \}, & 
l+m+n-2 \leq N, \\
0, & l+m+n-2 > N, 
\end{array}
\right.
\ee
where the generators $X$ are monomials of order $l \leq N$
\be
X_{i_1\cdots i_l} = x_{i_1}\cdots x_{i_l}. 
\ee
The case with $N = 1$ is precisely ${\cal A}_4$. 
As $N\rightarrow \infty$, this algebra approaches 
to a classical Nambu-Poisson structure 
on $C(\mathbb{R}^4)$. 

As the Nambu-Poisson algebra (\ref{classN}) is known 
to observe the fundamental identity, 
we only need to check that the truncation rule is 
compatible with it. 
Note that each term in the fundamental identity is of the form
$ [A_1, A_2, [A_3, A_4, A_5]] $. 
Let each $A_i$ to be a monomial of order $a_i$. 
Then this term is truncated to zero 
if $a_3+a_4+a_5-2 > N$ so that $[A_3, A_4, A_5]$ 
is truncated to zero, 
or if $a_1+\cdots+a_5-4 > N$ 
so that the outer bracket vanishes. 
However, since a monomial is at least of order 1, 
\footnote{
If one of the entries is of order 0 (that is, it is a constant), 
the Nambu bracket vanishes identically. 
}
we always have 
\be
a_1+a_2+a_3+a_4+a_5-4 \geq a_3+a_4+a_5-2, 
\ee
and hence the necessary and sufficient condition 
for truncation for every term in the fundamental 
identity is the same 
\be
\sum_{i=1}^5 a_i - 4 > N. 
\ee
Thus the fundamental identity is preserved by the truncation rule.

We can also try to define multiplication by 
truncating the products of monomials as 
\be \label{prod1}
X_{i_1\cdots i_l} \cdot X_{j_1\cdots j_m} = 
\left\{\begin{array}{ll}
X_{i_1\cdots i_l j_1\cdots j_m}, & l+m \leq N \\
0, & l+m> N.
\end{array}\right.
\ee
Again, one can check that the Leibniz rule, 
which is known to hold for the case $N = \infty$, 
is compatible with the truncation of products at finite $N$. 
Indeed, 
every term in the Leibniz rule condition 
\be
[A_1, A_2, A_3 A_4] = [A_1, A_2, A_3]A_4 + [A_1, A_2, A_4]A_3 
\ee
is truncated if and only if 
\be
a_1+a_2+a_3+a_4-2 > N. 
\ee

To define the metric, it is natural to use 
the integration over the underlying manifold. 
Decomposing the integration over the space of $x_i$ 
into the radial part and the integration over $S^3$, 
we define the metric as 
\be
\la A_1, A_2\ra = \int_{S^3} d^3\Omega \int_0^{\infty} dr \rho(r) A_1 \cdot A_2,
\ee
where we introduced a distribution $\rho(r)$ 
so that the integrals converge for polynomials of $x_i$. 
If we are considering the Nambu structure on 
a truncated set of functions on $S^3$ of radius $R$, 
we should take $\rho(r) = \delta(r - R)$.

Roughly speaking, treating $x_i$ as coordinates on $S^3$ 
is equivalent to imposing the constraint 
\be
\sum_{i=1}^4 x_i^2 = 1 
\ee
on the algebra of polynomials of $x_i$'s. 
Since $\sum_i x_i^2$ 
is a central element in the 3-algebra, i.e. 
\be \label{rad1} 
[\sum_i x_i^2, X_{i_1\cdots i_l}, X_{j_1\cdots j_m}] = 0, 
\ee
this constraint is consistent with the Nambu structure. 
However, the constraint is not compatible with 
the truncation rule for the Nambu bracket (\ref{Ntrunc}) 
or the product (\ref{prod1}). 
Thus we should not impose the constraint 
except when we compute the metric. 
The metric of $\la A, B\ra$ should be computed by 
first multiplying $A\cdot B$ with the truncation (\ref{prod1}), 
and then treating the product as a classical function on $S^3$ 
and integrate.

It is easy to see the the metric defined this way 
is not positive definite. 
Consider the norm of $A = x_1 - a x_1^{m}$, 
where $m$ is an odd number between $N/2+1$ and $N-1$. 
Its norm is 
\be
\la A, A\ra = \int_{S^3} x_1^2 - 2a \int_{S^3} x_1^{m+1},  
\ee 
where the term $\la x_1^m, x_1^m \ra$ is absent  
because $x_1^m\cdot x_1^m = 0$ according to (\ref{prod1}). 
While both terms on the right hand side are non-zero, 
one can choose $a$ to be sufficiently large so that 
the norm is negative.

\subsection{An Extension of ${\cal A}_4$}

An algebra with $4(N+1)$ generators 
$\{T^{(a)}_i: a = 0,\cdots, N, i = 1, \cdots, 4\}$ 
can be defined by 
\be \label{gradeN} 
[ T^{(a)}_i, T^{(b)}_j, T^{(c)}_k ]
=\left\{
\begin{array}{ll}
\epsilon_{ijkl} \; T^{(a+b+c)}_l, & a+b+c \leq N, \\
0, & a+b+c > N.
\end{array}
\right.
\ee

To check that the Nambu bracket (\ref{gradeN}) 
preserves the fundamental identity, 
we only need to check that the truncation rule 
is compatible with the fundamental identity, 
since this bracket is essentially just 
a grading of direct sums of ${\cal A}_4$. 
For a term in the fundamental identity 
\be
[T^{(a)}_i, T^{(b)}_j, [T^{(c)}_k, T^{(d)}_l, T^{(e)}_m]], 
\ee
we note that it is truncated if $c+d+e>N_2$ 
(so that the inner bracket is zero), 
or if $a+b+c+d+e > N$ 
(so that the outer bracket is zero). 
However, since $a, b \geq 0$, 
we always have 
$a+b+c+d+e > c+d+e$, 
and thus the necessary and sufficient condition for this term 
to be truncated to zero is just $a+b+c+d+e > N$. 
Since this condition is the same for all terms 
in the fundamental identity, 
the fundamental identity is preserved. 

One can further extend the 3-algebra form the linear space 
spanned by $T^{(a)}_i$'s to 
polynomials of the generators truncated at order $N$. 
Let 
\be
T^{(a)}_i T^{(b)}_j = 
\left\{\begin{array}{ll} 
T^{(b)}_j T^{(a)}_i, & a+b\leq N, \\
0, &  a+b> N. 
\end{array}\right.
\ee
The space of polynomials of $T^{(a)}_i$'s is thus spanned by 
the monomials 
$\{ T^{(a_1)}_{i_1}\cdots T^{(a_k)}_{i_k}: \sum_{r=1}^k a_r \leq N \}$. 
The Nambu bracket on this space can be defined 
by imposing the Leibniz rule
\be
[A^{(a)}, B^{(b)}, C^{(c)}D^{(d)}] = [A^{(a)}, B^{(b)}, C^{(c)}]D^{(d)}
+[A^{(a)}, B^{(b)}, D^{(d)}]C^{(c)}, 
\ee
where $A^{(a)}$ is a monomial $T^{(a_1)}_{i_1}\cdots T^{(a_k)}_{i_k}$ 
of level $\sum_{r=1}^k a_r = a$, etc. 
Note that the truncation rule of every term above 
is that each term vanishes if and only if $a+b+c+d \geq N$.

For a given function $f(a)$ with the property 
\be
f(a) = 0 \quad \mbox{for} \quad a > N,
\ee 
the invariant metric can be defined as
\be
\la T^{(a)}_i, T^{(b)}_j \ra
=
%\left\{
%\begin{array}{ll}
f(a+b) \delta_{ij} \qquad \mbox{for} \quad 
a, b = 0, \cdots, N, \;\; i, j = 1, \cdots, 4. 
%& a, b \leq N_2, \\
%0, & a \; \mbox{or} \; b > N.
%\end{array}\right.
\ee
Apparently all generators of level $a > N/2$ are null.

\subsection{Truncation of a Nambu-Poisson Algebra} 
\label{general1}

While Nambu-Poisson algebras are always Lie 3-algebras 
of infinite dimensions, 
it is sometimes possible to truncate the Nambu-Poisson algebra 
to a finite dimensional Lie 3-algebra. 
We have seen such an example in section \ref{truncateS3}. 
In fact, the same can be done for all linear Nambu-Poisson algebras. 
Starting with a linear Nambu-Poisson algebra, 
one can impose a truncation over monomials of 
the coordinates of order larger than $N$. 
The reason why this is a consistent truncation 
for the Nambu bracket 
is essentially the same as the arguments in section \ref{truncateS3}.

\subsection{Level Extension of a 3-Algebra}
\label{general2}

In the above we have seen that 
the notion of an additive level can be introduced 
to extend a given 3-algebra to a larger algebra. 
More precisely, given a 3-algebra 
\be
[T_i, T_j, T_k] = f_{ijk}{}^l T_l, 
\ee
with an invariant metric $h_{ij}$, 
we can define a new 3-algebra for generators $T^{(a)}_i$ 
($a =N_1, \cdots, N_2$ with $N_1\geq 0$) 
\be
[T^{(a)}_i, T^{(b)}_j, T^{(c)}_k] = f_{ijk}{}^l T^{(a+b+c)}_l. 
\ee
When $N_1 = 0$ the original 3-algebra is embedded at level $0$. 

A nontrivial choice of the metric is 
\be
\la T^{(a)}_i, T^{(b)}_j \ra = f(a+b) h_{ij}, 
\ee
for an arbitrary function $f(a)$ such that 
\be
f(a) = 0 \quad \mbox{for} \quad a > N_1+N_2. 
\ee
To check that this is invariant, 
we note that  
\be
\la [T^{(a)}_i, T^{(b)}_j, T^{(c)}_k], T^{(d)}_l \ra + 
\la T^{(c)}_k, [T^{(a)}_i, T^{(b)}_j, T^{(d)}_l] \ra  
= (f_{ijk}{}^m h_{ml} + f_{ijl}{}^m h_{mk}) f(a+b+c+d) 
= 0,  
\ee
whenever there is no truncation in both terms. 
When there is a truncation, we either have 
$a+b+c > N_2$ or $a+b+d > N_2$. 
This implies that $a+b+c+d > N_1+N_2$, 
and the equality above still holds 
because $f(a+b+c+d) = 0$. 

This is not the most general solution for the invariant metric. 
While generators $T^{(a)}_i$ at level $a<3N_1$ 
can never appear on the right hand side of a Nambu bracket, 
it is impossible to write down any constraint for 
the metric components 
$\la T^{(a)}_i, T^{(b)}_j \ra$ with $a, b < 3N_1$. 
Those components are thus arbitrary.

\subsection{A Conjecture}

The reason why examples of 3-algebra are so rare can be 
intuitively understood by noting the resemblance 
between the fundamental identity and the Pl\"{u}cker relation 
when a positive-definite metric is assumed. 
In the appendix we give a more detailed analysis of 
the fundamental identity with an effort to make its connection  
to the Pl\"{u}cker relation more manifest. 
We hope this will help us understand 
the fundamental identity better in the future. 
 
In \cite{MVV} it was conjectured that an $n$-algebra
is always a direct product of $n$-algebras of
dimension $n$ and $(n+1)$ and some trivial algebras.
This conjecture is ruled out by some of the examples listed above.
On the other hand, except ${\cal A}_4$ and the trivial algebra
(and their direct products), 
none of the examples we have so far
has a metric which is positive definite. 
All of them have generators of zero-norm.
Hence we conjecture that 
{\em all finite dimensional 3-algebras
with positive-definite metrics
are direct products of ${\cal A}_4$ with trivial algebras}. 
In other words, 
{\em except direct products of ${\cal A}_4$ with trivial algebras, 
all finite dimensional 3-algebras have generators of zero-norm}. 

A weaker form of the conjecture has already been studied 
in \cite{Gustavsson:2008dy}. 
There it was shown that nontrivial finite-dimensional 
generalization of ${\cal A}_4$, 
which is associated to the Lie algebra 
$SO(4) \simeq SU(2)\times SU(2)$, 
to other semi-simple Lie algebras is essentially impossible. 

For an algebra with a positive-definite metric,
we can always choose a new basis of generators such that
the metric is the identity matrix $\delta_{ab}$.
It follows from the invariance of the metric
\be
\langle [T_a, T_b, T_c], T_d \rangle + \langle T_c,[T_a, T_b, T_d] \rangle = 0
\ee
that
\be
f_{abcd} = - f_{abdc} \qquad (f_{abcd} \equiv f_{abc}{}^e h_{ed}). 
\ee
Since the structure constants are by definition skew-symmetric
with respect to the first 3 indices,
in this case
the 3-algebra structure constants are totally-antisymmetrized.
 
Assuming that the structure constants
are totally-antisymmetrized,
we checked using computers that all 3-algebras
with no more than 8 generators are either trivial
or are a direct product of the 4-generator algebra ${\cal A}_4$
with a trivial algebra.

The almost unavoidable appearance of 
the zero-norm (or null) generators is very interesting
from the viewpoint of physical applications.
For a dynamical variable $X$ living in the space of a 3-algebra
with generators $\{T_A\}$,
\be
X = X_A T_A,
\ee
its canonical kinetic term
\be
\langle \del_{\mu} X, \del_{\mu} X \rangle
\ee
there is no quadratic term for $X_A$ if $T_A$ is a null generator.
Hence the degrees of freedom associated with
the zero-norm generators are not dynamical.
They can be integrated out and
their equations of motion are constraints.
Therefore, each zero-norm generator
corresponds to a gauge symmetry.
Similarly, a negative norm generator corresponds to a ghost.

Infinite dimensional algebras with positive definite metrics
are easy to construct. 
As we mentioned in section \ref{sec:NP}, 
for any Nambu-Poisson structure on
the algebra $C({\cal M}_3)$ of functions on
a 3-dimensional space ${\cal M}_3$,
the Nambu-Poisson tensor field defines
a volume form on ${\cal M}_3$, 
which can be used to define an integral and then a metric. 
Whenever the volume form is everywhere non-vanishing, 
this metric is positive definite.

 %%%%%%%%%%%%%%%%%%%%%%%%%%%%%%%%%%%%%%%%%%%%%%%%%%%%%
\section{Representations of Nambu Bracket by Cubic Matrix}
\label{sec:representation}

\subsection{Motivation}

We would like to study representations of the Lie 3-algebra 
in this section. 
The first question is whether it is possible to 
represent the generators as matrices, 
which form an associative algebra. 
A natural definition of the quantum Nambu bracket is 
\cite{Nambu,Takhtajan} 
\be
[A, B, C] = ABC-ACB+BCA-BAC+CAB-CBA 
\ee
for an associative algebra with elements $A, B, C$. 
For the algebra ${\cal A}_4$, 
there are representations of arbitrary dimension $N\geq 2$ 
\cite{Kawamura:2003cw} based on 
the $N\times N$ irreducible representation of $su(2)$. 
Let $J^i$ ($i=1,2,3$) be the $N=2j+1$ dimensional 
irreducible representation of $su(2)$, 
then 
\be
R(T^i) = \frac{1}{(j(j+1))^{1/4}}J^i, \qquad
R(T^4) = (j(j+1))^{1/4} I,
\ee
where $i = 1,2,3$ and $I$ is the unit matrix, 
is a representation of ${\cal A}_4$. 
 
A problem with this representation is that 
the eigenvalues of $R(T^4)$ are fully degenerate. 
Interpreting $R(T^i)$ as some sort of quantum 
coordinates of $\mathbb{R}^4$, 
the geometric picture of this algebra is 
a fuzzy 2-sphere embedded in $\mathbb{R}^4$, 
with its 4-th coordinate fixed by 
\be
x^4 = (j(j+1))^{1/4}. 
\ee
On the other hand, in the physical applications 
we have in mind, 
one would like to interpret ${\cal A}_4$ as 
a fuzzy 3-sphere. 

Formally, ${\cal A}_4$ is a generalization of $su(2)$. 
While the adjoint representation of $su(2)$ is 
\be
(J_i)_{jk} = \epsilon_{ijk}, 
\ee
one is tempted to conjecture that for ${\cal A}_4$ 
we have a representation of the form 
\be
R(T^i)_{jkl} \sim \epsilon_{ijkl}. 
\ee
This is not exactly correct but 
we do have a representation of a similar form, 
which will be given below in (\ref{R1}). 
The point here is that although our lives 
would be much easier if we could just use matrices 
to represent Lie 3-algebras, 
but for the example of ${\cal A}_4$, 
it seems more appropriate to use objects 
with 3 indices.

There is also some physical motivation suggesting 
the use of cubic matrices. 
A long-standing puzzle about the low energy theory of 
coincident M5-branes is the following. 
In analogy with the case of D-branes, 
we imagine that cylindrical open membranes 
stretched between 2 M5-branes 
account for the low energy fields on M5-branes, 
and thus the low energy effective theory of $N$ M5-branes 
is expected to be a non-Abelian gauge theory 
with $N^2$ degrees of freedom. 
On the other hand, 
anomaly and entropy computations suggest that 
the M5-brane world-volume theory has $N^{3}$ degrees of freedom
\cite{N3}. 
Recently, arguments were presented based on 
considerations of membrane scattering amplitudes 
in the large $C$ limit, 
suggesting that the dominating configuration of 
membranes connecting M5-branes is not 
a cylindrical M2-brane stretched between 2 M5-branes, 
but rather a triangular M2-brane stretched among 3 M5-branes 
\cite{Ho:2007vk}. 
The low energy fields on M5-branes should 
hence appear as objects with 3 indices. 
As a supporting evidence, 
BPS configurations of membranes 
stretched among 3 M5-branes were found in \cite{Lee:2006gqa}. 
Therefore it is natural to introduce cubic matrices $X^{i}_{\alpha\beta\gamma}$, 
$i=1,2,3,4$ and $\alpha,\beta,\gamma=1,...,N$, 
to represent the spatial coordinates of open membranes 
with boundaries divided into 3 sections belonging to 
3 M5-branes ($\alpha\beta\gamma$).

\subsection{Realization by Cubic Matrices}

Cubic matrices were introduced in \cite{Kawamura:2002yz,Kawamura:2003cw}. 
A cubic matrix is an object with 3 cyclic indices 
\be
A_{ijk} = A_{jki} = A_{kij}.
\ee 
A triplet product of cubic matrices is defined as
\be \label{tripletm}
(A, B, C)_{ijk} = \sum_l A_{lij} B_{lki} C_{ljk}.
\ee
While Einstein's summation convention sums over indices repeated twice,
we will only sum over indices repeated thrice.\footnote{
Because of this property, this triplet product is not
invariant under the rotation (or the unitary transformation) of the indices.
It motivates us to introduce a generalized product in \S\ref{sec:irreducibility}.}
The Hermitian conjugation is defined by 
\be
A^{\dagger}_{ijk} = A^{\ast}_{kji},
\ee
and the inner product of two cubic matrices by 
\be
\langle A | B \rangle \equiv \sum_{ijk} A^*_{ijk} B_{ijk}.
\ee
Note that we used slightly different notations 
for the inner product for cubic matrices $\la \cdot | \cdot \ra$ 
and the inner product for 3-algebra $\la \cdot, \cdot \ra$. 

The cubic matrix algebra has some interesting properties. 
For example, it can be used to give a formulation 
of the generalized uncertainty relation for 3 observables 
\cite{Kawamura:2003cw}. 
The algebra of cubic matrix also naturally arises 
when we consider the scattering of open membranes 
in a large $C$ field background \cite{Ho:2007vk}. 

The Nambu bracket is defined for cubic matrices as
\be\label{e:Nambu}
[A, B, C] = (A, B, C) + (B, C, A) + (C, A, B) - (C, B, A) - (B, A, C) - (A, C, B). 
\ee

\subsection{Representations for ${\cal A}_4$}

The algebra ${\cal A}_4$ (\ref{algebra}) 
has been studied in the context of cubic matrices 
as the ``generalized spin algebra'' 
\cite{Kawamura:2003cw}. 

A $4\times 4\times 4$ representation of the algebra (\ref{algebra}) is
\be \label{R}
R(T^i)^{jkl} =
\left\{\begin{array}{ll}
e^{i\Omega^i_{jkl}} & \mbox{for} \quad i \neq j \neq k \neq l; \\
0,  & \mbox{ otherwise} .
\end{array}\right.
\ee
$\Omega^i_{jkl}$ is anti-symmetric 
$\Omega^i_{jkl}=-\Omega^i_{kjl}$, 
and cyclic
$\Omega^i_{jkl} = \Omega^i_{klj}$.
%We can imagine points i,j,k,l forms a tetrahedron, 
%and  $\Omega_{jkl}$ is $\frac{\pi}{8}$ 
%whenever loop(jkl) is oriented toward the interior of tetrahedron;
%on the other hand, $\Omega_{jkl}$ is $-\frac{\pi}{8}$.\\
% \begin{figure}
% \label{tetrahedron}
% \hskip-1.5cm
% \includegraphics[scale=0.4]{tetrahedron.pdf}
% \caption{tetrahedron}
% \end{figure}
They satisfy
\be
\Omega^i_{jkl} - \Omega^j_{kli} + \Omega^k_{lij} - \Omega^l_{ijk}
= \frac{\pi}{2} \epsilon_{ijkl}.
\ee
The sign of each term corresponds to the orientation 
of a face of a tetrahedron. %(see fig.\ref{tetrahedron}).
One way to assign values to $\Omega$'s is 
\be
\Omega^i_{jkl} = \frac{\pi}{8}\epsilon_{ijkl}.
\ee
In this case (\ref{R}) can be expressed as 
\be \label{R1}
R(T^i)^{jkl} = |\epsilon_{ijkl}| e^{i\epsilon_{ijkl}\pi/8}.
\ee
Obviously $R(T_i)$'s are all Hermitian. 

This representation $R$ has
\be \label{radius} 
\sum_{klm} R(T^k)_{lmi} R(T^k)_{lmj} = 3! \delta_{ij},
\ee
which can be viewed as the analogue of 
the condition 
\be
\sum_{i = 1}^4 X_i^2 = r^2
\ee
that defines a 3-sphere of radius $r$ in $\mathbb{R}^4$. 
Therefore it is natural to associate ${\cal A}_4$ to the notion of 
a fuzzy 3-sphere. 
Note that 
this algebra is different from 
the definition of fuzzy 3-sphere in \cite{Guralnik:2000pb}. 

%The equation above also implies that 
%\be
%\sum_i \langle R(T^i) | R(T^i) \rangle = \sum_{ijkl} R(T^i)_{jkl}R(T^i)_{jkl} = 4!.
%\ee

Representations of arbitrary dimension $N > 4$ 
can be found in \cite{Kawamura:2003cw}.

\subsection{Construction of Higher Representations}

Here we would like to discuss a question 
about cubic matrix representations for a generic Lie 3-algebras, 
that is, how to construct new representations 
from given representations. 
Like the representation by matrices, 
it is possible to construct higher dimensional representations 
by the direct sum and the direct product for
the representation by cubic matrices.

Suppose $R_i(T^a)$ ($i=1,2$) is an $N_i$ dimensional cubic matrix
which satisfies a given 3-algebra (not necessarily $\cA_4$). 
There are several systematic ways to construct new cubic matrix 
representations of the same 3-algebras from $R_i$: 
\begin{enumerate}
\item Direct sum representation $R_1\oplus R_2$ ($N_1+N_2$ dim):
\ba
&& (R_1\oplus R_2(T^a))_{ijk}\nn\\
&&~~~~~~=
\left\{
\begin{array}{ll}
R_1(T^a)_{ijk}\quad & \mbox{if } i,j,k\in \left\{
1,\cdots, N_1
\right\}, \\
R_2(T^a)_{i-N_1,j-N_1,k-N_1}\quad & \mbox{if } i,j,k\in \left\{
N_1+1,\cdots, N_1+N_2
\right\}, \\
0 & \mbox{otherwise}.
\end{array}
\right.
\ea

\item Direct product representations $R_1\otimes R_2$
which has dimension  $N_1N_2$:
\ba
(R_1\otimes R_2)_{IJK}=(R_1(T^a))_{ijk}\delta_{i'j'k'} \pm \delta_{ijk} (R_2(T^a))_{i'j'k'},
\qquad
\delta_{ijk}:=\delta_{ij}\delta_{ik}\,.
\ea
Here $I,J,K$ is the combination of two indices such as $I=(i,i')$,
$J=(j,j')$, $K=(k,k')$.  $i,j,k$ are in $1,\cdots, N_1$ and $i',j',k'$ are in $1,\cdots, N_2$.
We can take both sign in the second term since $-R_2(T^a)$ is also the
representation of the 3-algebra.

\item Tensor product $R(T^a)\otimes \cZ$ 
with constant cubic matrix $\cZ$ which satisfies
\ba\label{Z}
(\cZ,\cZ,\cZ)=\cZ.
\ea
If the size of $\cZ$ is $n\times n\times n$, the dimension of the representation
is $nN$.
There are many choices of $\cZ$.
Somewhat systematic construction of $\cZ$ is given later.

By taking the direct product of the fundamental representation of
$\cA_4$, one can obtain $4^n$ dimensional representations systematically.

In the representation theory of matrices, one may use the 
unitary transformation by which the representation matrix becomes
block diagonal form.  This notion, however, does not have straightforward
generalization to the cubic matrices.

\end{enumerate}

\paragraph{Construction of cubic projector $\cZ$}
%Eq.(\ref{Z}) seems to be similar to the projector equation  (\ref{cstr}).
Straightforward solutions of (\ref{Z}) are the diagonal cubic matrices,
\ba
Z_{ijk}=z_i\delta_{ijk},\qquad
z_i=\pm 1,0\,. 
\ea

For less trivial solutions, we observe that
eq.(\ref{Z}) resembles the projector equation. 
It motivates us seek solutions of the form,
\ba\label{vvv}
\cZ_{ijk}=v_i v_j v_k
\ea
where $v_i$ is an vector in $n$ dim space.

By requiring eq.(\ref{Z}), we obtain,
\ba
(\cZ,\cZ,\cZ)_{ijk}=(\sum_l v_l^3) v_i^2 v_j^2 v_k^2\,.
\ea
So if
\ba
v_i^2=(\sum_l v_l^3)^{-1/3} v_i, 
\ea
(\ref{vvv}) gives a solution to (\ref{Z}).
The general solution to this is 
\ba\label{v}
&& v_i=c \epsilon_j,\quad
\epsilon_j=\pm 1, 0, \\
&& c=\left( \sum_i \epsilon_i \right)^{-1/6} .
\ea

This construction can be generalized by using $r(<n)$ vectors 
$v^{(\alpha)}_i$ ($\alpha=1,\cdots,r$),
where each $v^{(\alpha)}$ takes the form (\ref{v}) and
the cubic orthogonality relation,
\ba
\sum_i v^{(\alpha)}_i v^{(\beta)}_i v^{(\gamma)}_i \propto
\delta_{\alpha\beta\gamma}.
\ea
Then,
\ba
&&\cZ_{ijk}=\sum_{\alpha=1}^r \cZ^{(\alpha)}_{ijk},\quad
\cZ^{(\alpha)}_{ijk}=
v^{(\alpha)}_iv^{(\alpha)}_j v^{(\alpha)}_k\\
&& (\cZ^{(\alpha)},\cZ^{(\beta)},\cZ^{(\gamma)})=\left\{
\begin{array}{ll}
\cZ^{(\alpha)} \quad & \mbox{if }\alpha=\beta=\gamma\\
0 & \mbox{otherwize}
\end{array}
\right.
\ea
satisfies (\ref{Z}). One might refer to such $\cZ$ as rank $r$ cubic projector.

We note that this construction does not give all the cubic projectors.
Even for the $2\times2\times2$ case,  
a direct algebraic computation by computer shows that
there are extra solutions which do not take this form

\subsection{Comments on Irreducibility}
\label{sec:irreducibility}

As mentioned earlier, the non-invariance of the triplet product (\ref{tripletm})
under the rotation of the indices forces us to introduce a generalization
of the product by using
a symmetric cubic  matrix $\cK$,
($\cK_{i_{\sigma(1)}i_{\sigma(2)}i_{\sigma(3)}}=\cK_{i_1i_2i_3}$),
\ba
% [A,B,C]&=& (A,B,C)+(B,C,A)+(C,A,B)\nn\\
% &&-(B,A,C)-(A,C,B)-(C,B,A)\\
(A,B,C)_{ijk}&=& \sum_{n,m,l,i',i'', j', j'', k', k''} \cK_{nml} A_{ni'j''}B_{mk'i''} C_{lj'k''}
\cK_{ii'i''}\cK_{jj'j''} \cK_{kk'k''}
\ea
where the indices $i,j,k,n$ run from 1 to $N$.
Usually we take $\cK_{ijk}=\delta_{ijk}$.  
% Here, we take
% more general form since we will make unitary (orthogonal)
% transformations which will be needed in the projection. 
We note that there is no orthogonal transformation
which keeps $\delta_{ijk}$ invariant.  In the general form above,
the summations are taken only for doubly repeated indices, so the notion
of the orthogonal transformation remains the same.

Suppose we consider a triplet product algebra such as
$[J^a, J^b, J^c]=i\epsilon_{a,b,c,d} J^d$, ($J^a:=R(T^a)$) and try to find
``irreducible decomposition''.  We introduce the orthogonal projectors
$p_{ij}$ and $q_{ij}$ which satisfy
\ba\label{cstr}
p^2=p,\quad
q^2=q,\quad
p^t=p,\quad
q^t=q,\quad
pq=0,\quad
p+q=1\,.
\ea
We note that such projector may be written as,
\ba\label{pq}
p=g\left(\begin{array}{cc} I_d&0\\ 0&0\end{array}\right)g^t\,,\quad
q=g\left(\begin{array}{cc} 0&0\\ 0&I_{N-d}\end{array}\right)g^t\,,\quad
g\in O(N,\mathbf{R})
\ea

One may define the algebra be reducible if there exists a pair
$p,q$ as above and they satisfy
\ba
&& \sum_{ij} (J^a)_{ijk} p_{ii'}q_{jj'} =
\sum_{jk} (J^a)_{ijk} p_{jj'}q_{kk'} =
\sum_{ij} (J^a)_{ijk} p_{kk'}q_{ii'} =
0\,,
\label{cond:J}
\\
% &&\sum_{ij} (\cR)_{ijk} p_{ii'}q_{jj'}= 
% \sum_{jk} (\cR)_{ijk} p_{jj'}q_{kk'}=
% \sum_{ki} (\cR)_{ijk} p_{kk'}q_{ii'}=0\,,
% \label{cond:R}
% \\
&&\sum_{ij} (\cK)_{ijk} p_{ii'}q_{jj'}= 
\sum_{jk} (\cK)_{ijk} p_{jj'}q_{kk'}=
\sum_{ki} (\cK)_{ijk} p_{kk'}q_{ii'}=0\,.
\label{cond:K}
\ea
If these identities are satisfied, 
we have a $d$ dimensional representation by redefining
the generators and the cubic product at the same time as 
\ba \label{Jtrans}
&&J^a\rightarrow  (\tilde J^a)_{ijk}=\sum_{i'j'k'} (J^a)_{i'ij' k'}p_{i'i}p_{j'j} p_{k'k}, \\
% &&\cR\rightarrow  (\tilde \cR)_{ijk}=\sum_{i'j'k'} (\cR)_{i'ij' k'}p_{i'i}p_{j'j} p_{k'k}
% \label{Rtrans}\\
&&\cK\rightarrow  (\tilde \cK)_{ijk}=\sum_{i'j'k'} (\cK)_{i'ij' k'}p_{i'i}p_{j'j} p_{k'k}.
\label{Ktrans}
\ea

% When $\cR_{ijk}=\cK_{ijk}=\delta_{ijk}$, the conditions (\ref{cond:R},\ref{cond:K})
% give,
% \ba
% p_{ki} q_{kj}=0
% \ea
% for any $k,i,j$. This condition is very restrictive. For given $k$, if one of
% $p_{ki}\neq 0$, one must impose $q_{kj}=q_{jk}=0$ for any $j$.
% It implies, after all,  that we $p,q$ becomes
%  $p=diag(\epsilon_1,\cdots,\epsilon_N)$ and
% $q=diag(1-\epsilon_1,\cdots,1-\epsilon_N)$ where $\epsilon_i=0$ or $=1$.

\paragraph{An example of reducible representation}

For a given representation $J^a$, 
the representation $\cJ^a=J^a\otimes \cZ$, where $\cZ_{\bi\bj\bk}$ is
written as (\ref{vvv}), gives an example of the reducible representation.
The projectors are,
\ba
p_{IJ} = \delta_{ij}\frac{v_{\bi} v_{\bj}}{\sqrt{|v|^2}}\,,
\quad
q_{IJ} = \delta_{ij}\left(1-\frac{v_{\bi} v_{\bj}}{\sqrt{|v|^2}}\right)\,.
\ea
In this sense, the tensor product with the cubic projector gives
a good example of the reducible representation in our sense.
We note, however, that the cubic matrices $\cK$ which defines
the cubic product is not given by the original
definition $\delta_{ijk}$ because of eq.(\ref{Ktrans}).

\paragraph{Failed example: (anti-)symmetrization}

In case of the Lie algebra, the tensor product of two fundamental
representations are reducible. Reduction to the irreducible representation
can be obtained by using (anti-)symmetrization of indices. In the following,
We will argue that this will not  be so simple for the cubic case.

We consider a direct product representation of two fundamental
representations,
\ba
J^a_{IJK}&=& J^a_{ijk}\delta_{\bi \bj \bk}+J^a_{\bi \bj \bk}\delta_{ijk}
\ea
and $\cK_{IJK}=\delta_{ijk}\delta_{\bi\bj\bk}$.
Here we use the multi-indices $I,J,K$ to represent $i,\bi$ and so on.

We define the projections to the symmetric and anti-symmetric part 
as
\ba
p_{IJ}=\frac{1}{2}\left(
\delta_{ij}\delta_{\bi \bj} +\delta_{i\bj} \delta_{\bi j}
\right),\quad
q_{IJ}=\frac{1}{2}\left(
\delta_{ij}\delta_{\bi \bj} -\delta_{i\bj} \delta_{\bi j}
\right).
\ea
It is easy to see that $p,q$ satisfy the constraint (\ref{cstr}).
On the other hand, conditions (\ref{cond:J}--\ref{cond:K}) become
\ba
\sum_{IJ} J^a_{IJK}p_{IL} q_{JM}&=&\frac{1}{4}\left(
J^a_{lmk}\delta_{\bl\bm\bk}
-J^a_{l\bm k}\delta_{\bl m\bk}
+J^a_{\bl mk}\delta_{l\bm\bk}
-J^a_{\bl \bm k}\delta_{lm\bk}
\right.
\nn\\
&&\left.
+\delta_{lmk}J^a_{\bl\bm\bk}
-\delta_{l\bm k}J^a_{\bl m\bk}
+\delta_{\bl mk}J^a_{l\bm\bk}
-\delta_{\bl \bm k}J^a_{lm\bk}
\right)\\
\sum_{IJ} \delta_{IJK}p_{IL} q_{JM}&=&\frac{1}{2}\left(
\delta_{lmk}\delta_{\bl\bm\bk}
-\delta_{l\bm k}\delta_{\bl m\bk}
+\delta_{\bl mk}\delta_{l\bm\bk}
-\delta_{\bl \bm k}\delta_{lm\bk}
\right)
\ea
They do not vanish. It implies that the (anti-)symmetrization which works
in the construction of the representation of Lie algebra does not
work for cubic matrices.

\section{Application to Multiple M2-Branes}
\label{sec:application}

\subsection{Basu-Harvey Equation}
\label{sec:BH}

Generalizing Nahm's equation, 
which was used to describe the analogous 
configuration of D1-branes ending on D3-branes, 
the Basu-Harvey equation was proposed \cite{Basu:2004ed} 
to describe multiple M2-branes ending on an M5-brane 
\be \label{BH}
\frac{dX^i}{ds} +i \frac{K}{3!} \eps^{ijkl} [X^j, X^k, X^l] = 0, 
\ee
where $X^i(s)$'s represent spatial fluctuations of the M2-branes, 
and $s$ is a worldvolume coordinate.  
This equation admits a funnel solution:
\bea \label{ansatz}
X^i(s) &=& f(s) R(T^i),\\
f(s) &=& \frac{1}{\sqrt{2Ks}}, \label{fs}
\eea
where $T^i$ satisfies the $SO(4)$-invariant algebra ${\cal A}_4$ 
\be \label{algebra1}
[T^i, T^j, T^k] = i\eps^{ijkl} T^l, \qquad (i,j,k,l=1,2,3,4,)
\ee
and $R(T^i)$ is any representation of this algebra.

As we will see below, 
the Basu-Harvey equation can be interpreted 
as a BPS condition for the multiple M2-brane 
action of Bagger and Lambert \cite{Bagger:2007vi}, 
although it was first proposed without an underlying Lagrangian. 
On the other hand, this particular solution 
happens to define a Lie 3-algebra structure. 
It is possible to proceed for our present purpose 
without assuming a particular M2-brane action. 
% In fact, the energy (\ref{Energy}) proposed in \cite{Basu:2004ed}
% is different from the energy for 
% the M2-brane action of Bagger and Lambert \cite{Bagger:2007vi}. 
% (See section \ref{sec:BL}.) 

In order to give a proper geometrical interpretation to this solution, 
we also need to assume that the algebra (\ref{algebra1}) of $T^i$ 
describes a {\em fuzzy three-sphere} with radius $r$ given by 
\be
r^2 \equiv \sum_i (X^i)^2 \propto f^2(s) \propto \frac{1}{Ks}.
\ee
Hence
\be \label{ras}
r^2 = \frac{\alpha}{Ks}
\ee
for some constant $\alpha$.
The $T^i$'s then represent the Cartesian coordinates 
of the fuzzy 3-sphere. 
Furthremore, infinitesimal $SO(4)$ rotations are generated by 
\be
\d T^k = \Lambda_{ij} [ T^i, T^j, T^k ],  
\ee 
and the invariant metric is 
\be \label{inner1}
\langle T^i, T^j \rangle = \delta^{ij}. 
\ee

The energy proposed in \cite{Basu:2004ed} is 
\be \label{Energy}
E = T_2 N \int d^2\s \left[
a^2 \left|\frac{dX^i}{ds} -i \frac{K}{3!} \eps^{ijkl}[X^j, X^k, X^l]\right|^2
+ \left(1 +i \frac{C}{3!} \eps^{ijkl}\left\langle \frac{dX^i}{ds}\Big|[X^j, X^k, X^l]\right\rangle\right)^2
\right]^{1/2}, 
\ee
where $|A|^2 \equiv \la A|A\ra$.
We will specify the two constant parameters $a$ and $C$ below.  
%For the particular solution (\ref{ansatz}), 
%we naturally use the metric
%\be
%( T^i T^j ) = \delta^{ij} 
%\ee
%which is invariant
%\be
%( [T^i, T^j, T^k]T^l) + (T^k[T^i, T^j, T^l]) = 0.
%\ee

For $X^i = 0$ (or more generally when $\frac{dX^i}{ds} = 0 = [X^j, X^k , X^l]$),
the energy is that of $N$ D2-branes at rest: 
$E = T_2 N$ times the M2-brane volume.
The form of the energy $E$ is such that 
the Basu-Harvey equation (\ref{BH})
is a BPS condition.
One should choose $a$ as
\be
a^2 = \frac{C}{K}
\ee
so that the cross-term proportional to
$\langle \frac{dX^i}{ds} | [X^j, X^k, X^l] \rangle$ cancels in (\ref{Energy}),
otherwise the theory is not covariant.

For the funnel solution (\ref{ansatz}) and (\ref{fs}),
the energy is 
\be
E = T_2 N \int d^2 \s \left|
1+\frac{C}{K} \left\langle \frac{dX^i}{ds} \Big| \frac{dX^i}{ds} \right\rangle
\right|
= T_2 N L \int ds + T_2 N L \int ds \frac{C}{8K^2 s^3} \langle R(G^i) | R(G^i)\rangle.
\ee
According to (\ref{ras}),
\be
\int_0^{\infty} \frac{ds}{s^3} = \frac{2K^2}{\a^2} \int_0^{\infty} dr \; r^3,
\ee
and thus
\be \label{E/L}
E/L = T_2 N \int ds + \beta \int dr \; r^3,
\ee
where
\be \label{beta}
\beta = 2 T_2 N \frac{\langle R(G^i) | R(G^i)\rangle}{8\a^2} C.
\ee
We should choose $C$ such that
\be
\beta = 2\pi^2 T_5, 
\ee
where $T_5$ is the M5-brane tension. 
%and our convention is
%\be
%T_2 = \frac{M_{11}^3}{(2\pi)^2}, \qquad T_5 = \frac{M_{11}^6}{(2\pi)^5}.
%\ee

The derivation above goes through without the need of
a representation for the bracket in (\ref{BH}).
While the constant $C$ can be tuned to give the correct answer,
the needed $r^3$ dependence of the 2nd term in $E$
is also guaranteed by the relation (\ref{ras})
\be
r^2 \propto \frac{1}{s},
\ee
which is a direct result of the fact that the two terms in
the Basu-Harvey equation differ in the order of $X$ by 2.

After choosing $C$ properly to get the correct expression of energy
for the M2-M5 system,
$K$ is still a free parameter.
But we can always scale $X$ so that $K = 1$.

In the original work of Basu and Harvey \cite{Basu:2004ed}, 
they considered the fuzzy 3-sphere defined in \cite{Guralnik:2000pb}. 
What we have shown above is that actually the success of 
Basu-Harvey equation does not rely on a particular choice 
of how the fuzzy 3-sphere algebra (\ref{algebra1}) is realized. 
All we need are the general properties of the Lie 3-algebra.

\subsection{Multiple M2-Brane Action}
\label{sec:BL}

Bagger and Lambert \cite{Bagger:2006sk,Bagger:2007jr,Bagger:2007vi} proposed 
a supersymmetric Lagrangian for M2-branes  
for a given 3-algebra as 
\be
{\cal L} = -\frac{1}{2} \la D^{\mu}X^I, D_{\mu} X^I\ra 
+ \frac{i}{2} \la\bar\Psi, \Gamma^{\mu}D_{\mu}\Psi\ra 
+\frac{i}{4} \la\bar\Psi, \Gamma_{IJ} [X^I, X^J, \Psi]\ra 
-V(X) + {\cal L}_{CS},
\ee 
where $D_{\mu}$ is the covariant derivative, 
$V(X)$ is the potential term defined by 
\be
V(X) = \frac{1}{12}\la [X^I, X^J, X^K], [X^I, X^J, X^K]\ra, 
\ee
and the Chern-Simons action for the gauge potential is 
\be
{\cal L}_{CS} = \frac{1}{2}\epsilon^{\mu\nu\lam}
\left(f^{abcd}A_{\mu ab}\del_{\nu}A_{\lam cd} 
+ \frac{2}{3} f^{cda}{}_g f^{efgb} A_{\mu ab} A_{\nu cd} A_{\lam ef} \right). 
\ee
The SUSY transformation is defined by
\bea
\d X^I_a &=& i\bar{\eps}\Gamma^I \Psi_a, \\
\d \Psi_a &=& D_{\mu}X^I_a \Gamma^\mu\Gamma^I \eps  
- \frac{1}{6} X^I_b X^J_c X^K_d f^{bcd}{}_a \Gamma^{IJK}\eps, \\
\d \tilde{A}_{\mu}{}^b{}_a &=& 
i\bar{\eps}\Gamma_{\mu}\Gamma_I X^I_c \Psi_d f^{cdb}{}_a. 
\eea

While the fundamental identity is needed for the gauge symmetry 
of the multiple M2-brane theory, 
the invariant metric is also necessary to 
write down the gauge-invariant Lagrangian. 

For the background with $\Psi = \tilde{A} = 0$, 
a BPS condition should guarantee that 
\be
\left(\del_{\mu} X^I \Gamma^{\mu} \Gamma^I 
- \frac{1}{6}[X^I, X^J, X^K]\Gamma^{IJK}\right)\eps = 0
\ee
for some constant spinor $\eps$.
Assuming that $\del_t = \del_\s = 0$, 
for the constant spinor satisfying 
\be
\left(1 + \frac{i}{K} \Gamma^s\Gamma^{1234}\right) \eps = 0,
\ee
the BPS condition is guaranteed if 
\be \label{BH1}
\frac{dX^i}{ds} +i \frac{K}{3!} \eps^{ijkl} [X^j, X^k, X^l] = 0, 
\ee
where the superscript $s$ on $\Gamma^s$ 
denotes the direction in which $X^s$ is identified with 
the M2-brane worldvolume coordinate $s$, 
and $\Gamma^{1234} \equiv \Gamma^1 \Gamma^2 \Gamma^3 \Gamma^4$, 
and we also assumed that $X^I = 0$ except for $I = 1,2,3,4$.
We see that the Basu-Harvey equation is indeed a BPS condition for this theory 
if $K = \pm 1$ (this can always be achieved by scaling $X$). 

For a solution of the Basu-Harvey equation, 
the Hamiltonian density of the Bagger-Lambert model is simply 
\be
{\cal H} = \la \del_s X^I, \del_s X^I \ra. 
\ee
This coincides with the Hamiltonian proposed in \cite{Basu:2004ed}
up to a constant shift and overall factor.

Although the the connection between the Basu-Harvey equation 
and the Bagger-Lambert model begins to be clarified
we have an impression that there still remain some mysteries
which should be clarified in the future.
Incidentally, apart from the Basu-Harvey equation, 
the study of Bagger-Lambert model with boundaries \cite{Berman:2008be} 
is another approach to M5-branes from the M2-brane viewpoint.

\section{Comments}
\label{sec:comments}

\subsection{Lie 3-Algebra}

In this paper we discussed quite a few new examples of Lie 3-algebra 
of finite dimensions. 
Yet we still have the basic problem of lacking any mathematical structure 
analogous to the matrix algebra, 
which guarantees that the commutator defines a Lie algebra. 
The fundamental identity appears to be much more restrictive 
than the Jacobi identity, 
and we do not know much about how to solve it. 

The truncation of a Nambu-Poisson bracket 
(sections \ref{truncateS3}, \ref{general1}) can be used to 
construct a finite dimensional Lie 3-algebra. 
While the naive truncation works well, 
it will be desirable to find a deformed truncation 
such that the final 3-algebra possesses better properties. 
A possible motivation is to avoid 
negative norm generators in the algebra. 
Another example is that, 
for the truncated Nambu bracket on $S^3$, 
the radius constraint $x_i^2 = r^2$ can not be imposed 
until computing the metric. 
Although the linear dependence among functions 
will be fixed by the metric, 
and thus this will only result in some redundancy 
of the generators, 
similar to what happens when we use an over-complete basis 
of functions on a manifold, 
it would be better if this 3-algebra can be deformed 
such that the constraint can be imposed directly on the generators. 

One can apply the general procedures of section \ref{general2} 
to a given 3-algebra for an arbitrary number of times 
to obtain more and more new examples of Lie 3-algebras. 
Yet it remains to be seen how nontrivial these examples will be. 

For physical applications to multiple M2-branes, 
since we want the M2-branes turn into D2-branes 
upon compactifying a spatial direction, 
we hope to associate the $su(N)$ Lie algebra 
with a Lie 3-algebra for each $N$. 
So far we only know that ${\cal A}_2$ 
is associated with $su(2)$ \cite{Mukhi}. 
In section \ref{onegenerator}, 
we present a 3-algebra based on an arbitrary Lie algebra. 
However its metric is almost trivial. 
It is most desirable to find Lie 3-algebras 
associated to all $su(N)$'s.

\subsection{Cubic Matrices}

There are a few issues regarding cubic matrices 
which should be studied further in the future.

First, in the construction of higher representations,  we introduced
the direct product. In case of Lie algebra, such a procedure produces
{\em reducible} representations and we have to decompose them to
extract the irreducible representations.  In order to do similar reduction,
we need to define the corresponding notions of the direct sum representations
and the unitary equivalence between representations, i.e., 
representations $R$ and $R'$ are equivalent if there exists a unitary matrix 
$U$ such that $R'(T)=UR(T)U^\dagger$. 
For cubic-matrix representations, it is trivial to see that the direct sum gives a new
representation.  On the other hand, in order to define the unitary equivalence,
it is natural to use the Nambu bracket $\delta R=[R, K_1, K_2]$, 
for some $K_1$ and $K_2$, 
and we need to impose the fundamental identity 
in order to preserve the algebraic structure.  
However, the fundamental identity is not satisfied 
for generic elements of the cubic matrices. 
The subset of cubic matrices which is known to satisfy the
fundamental identity is the set of objects called ``normal matrices" 
\cite{Kawamura:2002yz}.
They are, however, an analogue of diagonal matrices and give rise to a trivial
change of the representation.  

Second, in this paper, we introduce only the triplet multiplication (\ref{tripletm}).
By composing it, we can generate functions of odd power.
This is not sufficient to produce all functions
on a fuzzy space to guarantee a proper classical limit. 
In order to generate a generic function,
we would need other type of products. 
As we commented in our previous paper \cite{Ho:2007vk},
for such a direction, it will be necessary to
introduce objects with more indices $\Psi_{i_1\cdots i_n}$.
How to construct a series of the products consistently remains 
a big challenge.

\subsection{Multiple M2-Branes}

Recently there is a very interesting paper \cite{Mukhi} 
which proposed a novel Higgs mechanism for 
the Bagger-Lambert model \cite{Bagger:2006sk} so that 
the multiple M2-brane action reduces to the D2-brane effective action 
upon compactification of a spatial coordinate. 
Later it was realized that 
\cite{VanRaamsdonk:2008ft,Lambert:2008et,Distler:2008mk} 
the moduli space for the model with the ${\cal A}_4$ algebra 
does not match with the moduli space for 2 M2-branes 
in flat space, 
but rather it matches with the moduli space for an orbifold. 
While this is a success of the Bagger-Lambert model, 
it is now even more urgent to consider more examples of 3-algebras 
for the Bagger-Lambert model to go beyond a single special case. 
It will be very interesting to see whether some of 
the examples provided in this work will correspond to 
a certain physical background for M2-branes in M theory. 
It will also be very intriguing to find out the physical 
interpretation of the ubiquitous zero-norm generators. 
In some of the examples there are also 
negative norm generators, 
which can potentially result in ghosts in the model. 
Perhaps those algebras with negative norm generators 
should be dismissed in certain applications, 
just like we usually avoid non-compact Lie groups 
in certain physical problems. 
It will be interesting to see whether 
there are other physical applications of 
the Lie 3-algebra besides M2-branes physics.

\section*{Note added}
After we submitted this paper to arXiv, we are informed that
the relation between the fundamental identity and the Pl\"ucker relation
was studied in \cite{FigueroaO'Farrill:2002xg} where a systematic study
fundamental identity in $\mathcal{D}=5,6,7,8$ was also carried out.

\section*{Acknowledgment}

We appreciate partial financial support from
Japan-Taiwan Joint Research Program
provided by Interchange Association (Japan)
by which this collaboration is made possible.

The authors thank Hidetoshi Awata, David Berman, Kazuyuki Furuuchi, Takeo Inami, 
Hsien-chung Kao, Xue-Yen Lin, Darren Sheng-Yu Shih, 
and Wen-Yu Wen for helpful discussions. 
P.-M. H. is supported in part by
the National Science Council,
and the National Center for Theoretical Sciences, Taiwan, R.O.C.
Y. M. is partially supported by
Grant-in-Aid (\#16540232) from the Japan
Ministry of Education, Culture, Sports,
Science and Technology.

\appendix

\section*{A. Relation with Pl\"ucker Relation}

Here we show that there is a direct relation between the
fundamental identity and Pl\"{u}cker relation which 
characterizes the locus of the Grassmannian manifold.
This bilinear relation appeared in a variety of context
in the physical literature, such as the exactly solvable
system (KP hierarchy etc.), 
free fermions on Riemann surface, topological string,
matrix model and so on \cite{plucker}.  Although this relation itself
is not new in mathematical literature (see for example \cite{Vaisman}),
it might shed a new light in the study of the fundamental identity 
(\ref{fip}).

To see the relation, we rewrite the structure constant by the metric,
by lowering the upper index by the metric,
$f_{a_1,\cdots,a_{p+1}}={f_{a_1,\cdots, a_p}}^{b} h_{b a_{p+1}}$,
 which gives the
rank $p+1$ anti-symmetric tensor.
It can be identified as the coefficients of the
$p+1$ vector by writing them with the wedge product of
the orthonormal basis of $n$ dimensional vector space $\e_1,\cdots, \e_n$,
\ba
|f\rangle = \sum_{a_1,\cdots, a_{p+1}} {f_{a_1\cdots a_{p+1}}} 
\e_{a_1}\wedge\cdots \wedge \e_{a_{p+1}}\,.
\ea
Pl\"ucker relation is a condition on the coefficient ${f_{a_1\cdots a_{p+1}}}$
when the $(p+1)$ vector $|f\rangle$ is written in the form,
\ba\label{product}
|f\rangle = \vv_1\wedge\cdots \wedge \vv_{p+1}
\,,\qquad
\vv_a \in \mathbf{R}^{n}\,.
\ea
The requirement is given by a set of bilinear relations,
\ba
\sum_{k=1}^{p+2} (-1)^k f_{a_1,\cdots, a_p, b_k}
f_{b_1,\cdots, b_{k-1},b_{k+1},\cdots, b_{p+2}}=0
\ea
where $(a_1,\cdots, a_p)$ and $(b_1,\cdots, b_{p+2})$ is the arbitrary 
number in $1,\cdots, n$.  The fundamental identity is obtained from
Pl\"ucker relation by putting $a_1=b_1=a$ and  take the sum over $a$.
Because of this procedure, the fundamental identity is a weaker condition
than the Pl\"ucker relation.

In particular, when
\ba
f_{a_1,\cdots, a_{p+1}}=\left\{
\begin{array}{ll}
\epsilon_{a_1,\cdots, a_{p+1}} \qquad & a_1,\cdots, 
a_{p+1}\in \left\{1,\cdots, p+1\right\}\\
0 & \mbox{otherwise}
\end{array}
\right.
\ea
the $(p+1)$-vector becomes
$|f\rangle =\e_1\wedge \cdots \wedge \e_{p+1}$.  Therefore, it satisfies
the Pl\"ucker relation and the fundamental identity.
We note that the direct sum of this $p$-algebra corresponds to
the $p$ vector of the form $\e_1\wedge \cdots \wedge \e_{p+1}+\e_{p+2}\wedge
\cdots \wedge \e_{2p+2}+\cdots$ which is definitely not of the form (\ref{product}). 
In this sense, the fundamental identity
allows a broader set of solutions than the Pl\"ucker relation.

In the application to the physics, it may be useful to rewrite these
relations by free fermions. To define them,
we consider space of $p$-vectors, $\cH_p$, ($p=0,1,\cdots,n$) where base
is spanned by exterior product of the basis,
$\e_{i_1}\wedge\cdots \wedge \e_{i_p}$, 
($ i_1 <\cdots <i_p$).
On this $p$-vector space, we introduce ``fermion" operators
$\psi_i,\psib_i$ ($i=1,\cdots,n$) as
\ba
&&\psib_a (\e_{i_1}\wedge\cdots \wedge \e_{i_p})
= \e_a\wedge \e_{i_1}\wedge\cdots \wedge \e_{i_p}\,,\\
&& \psi_a (\e_{i_1}\wedge\cdots \wedge \e_{i_p})
= \sum_{k=1}^{p} (-1)^{k-1}\delta_{a i_k}  
\e_{i_1}\wedge\cdots \wedge \e_{i_{k-1}}\wedge\e_{i_{k+1}}\cdots\wedge \e_{i_p}
\,.
\ea
These operators satisfy standard anticommutation relations,
\ba
\left\{\psi_i, \psib_j \right\}=\delta_{ij}\,,\quad
\left\{\psi_i, \psi_j \right\}=\left\{\psib_i, \psib_j \right\}=0\,.
\ea
The Pl\"ucker relation and the fundamental identity is then
written in terms of the fermions as,
\ba
&&\mbox{Pl\"ucker relation}:\quad
\sum_{i=1}^n \psi_i|f\rangle\otimes \psib_i |f\rangle=0\,,
\\
&& \mbox{Fundamental identity}:\quad
\sum_{i,j=1}^n \psi_j \psi_i|f\rangle\otimes \psi_j \psib_i |f\rangle=0\,.
\ea

\end{document}